\pdfoutput=1
\documentclass[prx,twocolumn,floatfix,superscriptaddress]{revtex4-1}
\usepackage{graphicx}
\usepackage{amsmath}
\usepackage{amssymb}
\usepackage[colorlinks=true, linkcolor=blue, urlcolor=blue,citecolor=blue]{hyperref}
\usepackage{bm}
\usepackage{braket}

\begin{document}
\title{Magnon versus electron mediated spin-transfer torque exerted by spin currents across antiferromagnetic insulator to switch magnetization of adjacent ferromagnetic metal}

\author{Abhin Suresh}
\affiliation{Department of Physics and Astronomy, University of Delaware, Newark DE 19716, USA}

\author{Marko D. Petrovi\'{c}}
\affiliation{Department of Physics and Astronomy, University of Delaware, Newark DE 19716, USA}

\author{Utkarsh Bajpai}
\affiliation{Department of Physics and Astronomy, University of Delaware, Newark DE 19716, USA}

\author{Hyunsoo Yang}
\affiliation{Department of Electrical and Computer Engineering, National University of Singapore, 117576, Singapore}

\author{Branislav K.~Nikoli\'c}
\email{bnikolic@udel.edu}
\affiliation{Department of Physics and Astronomy, University of Delaware, Newark DE 19716, USA}
\affiliation{Kavli Institute for Theoretical Physics, University of California, Santa Barbara, CA 93106-4030}

\begin{abstract}
The recent experiment [Y. Wang {\em et al.}, Science {\bf 366}, 1125 (2019)] on {\em magnon-mediated} spin-transfer torque (MSTT) was interpreted in terms of 
a picture where magnons are excited within an antiferromagnetic insulator (AFI), by applying nonequilibrium electronic spin density at  one of its surfaces,  so that their propagation across AFI deprived of conduction electrons eventually leads to reversal of magnetization of a ferromagnetic metal (FM) attached to the opposite surface of AFI.  However, microscopic (i.e., Hamiltonian-based) understanding of how magnonic and electronic spin currents, both of which can exert torque on localized magnetic moments within FM, are generated and interconverted at {\em multiple} junction interfaces is lacking. We employ a recently developed  time-dependent nonequilibrium Green functions combined with the Landau-Lifshitz-Gilbert equation (TDNEGF+LLG) formalism to evolve conduction electrons quantum-mechanically while they interact via self-consistent back-action with localized magnetic moments described classically by atomistic spin dynamics solving a system of LLG equations. Upon injection of square current pulse as the initial condition, TDNEGF+LLG simulations of FM-polarizer/AFI/FM-analyzer junctions show that reversal of localized magnetic moments within FM-analyzer is less efficient, in the sense of requiring larger pulse height and its longer duration, than conventional electron-mediated STT (ESTT) driving magnetization switching in standard FM-polarizer/normal-metal/FM-analyzer spin valve. Since {\em both} electronic, generated by spin pumping from AFI, and magnonic, generated by direct transmission from AFI, spin currents are injected into the FM-analyzer, its localized magnetic moments will experience combined MSTT and ESTT. Nevertheless,   by artificially turning off ESTT we demonstrate that MSTT plays a {\em dominant} role whose understanding, therefore, paves the way for all-magnon-driven magnetization switching devices with no electronic parts. 
\end{abstract}

\maketitle

\section{Introduction} \label{sec:into}

The magnon-mediated spin transfer torque (MSTT)~\cite{Yan2011,Hinzke2011,Kovalev2012,Cheng2018a,Cheng2018,Cheng2019} is a phenomenon where spin current carried by spin wave (SW) within an insulating or metallic magnetic material transfers spin angular momentum to its localized magnetic moments (LMMs). In the semiclassical picture~\cite{Kim2010,Evans2014}, SW is a  disturbance in the local magnetic ordering of a magnetic material in which LMMs precess around the easy axis with the phase of precession of adjacent moments varying harmonically in space over the  wavelength $\lambda$. The quanta of energy of SW behave as a quasiparticle, termed magnon, which carries energy $\hbar \omega$ and spin $\hbar$. The frequency $\omega$ of the precession is commonly in the GHz range of microwaves, but it can reach THz range in antiferromagnets~\cite{Jungfleisch2018,Baltz2018,Jungwirth2016,Zelezny2018}. 

The SWs are inevitably excited at finite temperature  as incoherent thermal fluctuations. But they can also be induced in controllable fashion by using external fields~\cite{Demidov2007}, or by injecting spin-polarized or pure spin currents~\cite{Demidov2016}, thereby leading to coherent propagation of SWs as a dispersive signal.  Since both electrons and magnons carry intrinsic angular momentum, their translational flow is equivalent  to a flux of spin angular momentum which is denoted as electronic and  magnonic (or SW) spin current~\cite{Yan2011,Chumak2015,Suresh2020}, respectively.

The MSTT provides an alternative to conventional electron-mediated spin-transfer torque (ESTT) where  electronic spin current transfers spin angular momentum to LMMs, on the proviso that electronic spin polarization is {\em noncollinear} to the direction of LMMs~\cite{Ralph2008,Tatara2019,Tsoi2000,Katine2000,Stiles2002,Wang2008b}. Since SWs in magnetic insulators can transmit spin current over $\sim \mu$m distances in the absence of Joule heating, all-magnon-driven magnetization dynamics and switching, without any electronic motion, has been envisioned~\cite{Chumak2015}. This requires, e.g.,  temperature gradients to excite SWs~\cite{Hinzke2011,Kovalev2012,Cheng2018a,Cheng2018} and the corresponding magnonic spin current. Another proposal~\cite{Cheng2019} is to insert a magnetic insulator barrier as a spacer between two ferromagnetic metals (FM) forming a magnetic tunnel junction where MSTT, driven by  asymmetric heating of two FM layers,  could  enhance conventional ESTT. Besides fundamental interest, MSTT-based devices are envisioned as ultralow dissipation platform for magnon-based memory, logic and logic-in-memory~\cite{Chumak2015}. 

\begin{figure}
	\includegraphics[width =\linewidth]{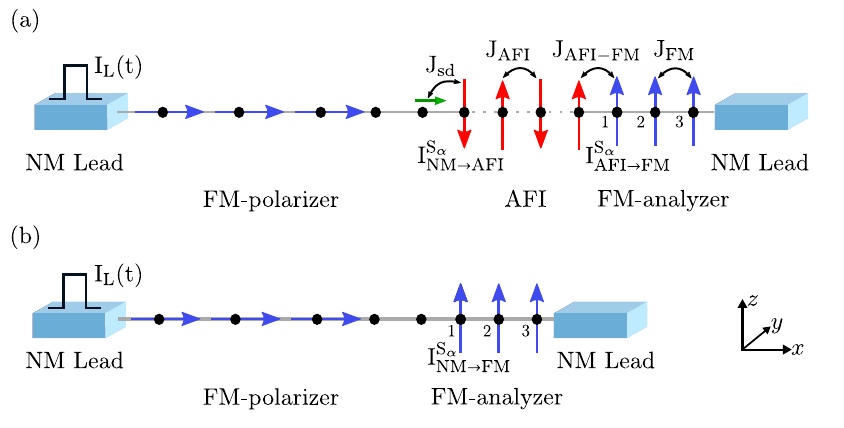}
	\caption{Schematic view of (a) FM/NM/AFI/FM and (b) FM/NM/FM two-terminal junctions whose active region (modeled as 1D tight-binding chain) is attached via semi-infinite NM leads (modeled as 1D tight-binding chains without any spin-dependent interactions) to two macroscopic reservoirs of electrons. The active region hosts: (a) three fixed LMMs (blue arrows) pointing along the $x$-axis which comprise FM-polarizer layer, followed by two sites without LMMs, followed by N\'{e}el-ordered 20 LMMs (red arrows) comprising AFI layer, and three free LMMs (blue arrows labeled 1--3) comprising FM-analyzer which receives MSTT and/or ESTT; (b) same as (a) but without AFI layer. A square voltage pulse  is applied (at \mbox{$t_0=1$ ps}) to inject unpolarized charge current $I_\mathrm{L}(t)$ from the left NM lead.  Also denoted are local spin currents: $I^{S_\alpha}_\mathrm{NM \rightarrow AFI}$ impinging on the first LMM of AFI in (a); $I^{S_\alpha}_\mathrm{AFI \rightarrow FM}$ impinging on the first LMM of FM-analyzer in (a); and $I^{S_\alpha}_\mathrm{NM \rightarrow FM}$ impinging on the first LMM of FM-analyzer in (b).}
	\label{fig:fig1}
\end{figure}

Following theoretical predictions~\cite{Yan2011}, a very recent experiment~\cite{Han2019} has  demonstrated MSTT-driven motion of magnetic domain wall in FM multilayer films based on  Co/Ni. Another experiment~\cite{Wang2019} has shown how SW excited in an antiferromagnetic insulator (AFI) NiO by spin-orbit torque~\cite{Manchon2019} from metallic surface of three-dimensional topological insulator (TI) Bi$_2$Se$_3$ was able to switch the magnetization of Py as FM-analyzer layer within Bi$_2$Se$_3$/NiO/Py heterostructure. Within AFI electrons do not move, hence  SWs are the sole carrier of spin currents. Increasing the thickness of the AFI layer improves its antiferromagnetic ordering, so that MSTT acting on the FM-analyzer Py reaches  an optimal magnitude at $\simeq 25$~nm thick NiO layer in Ref.~\cite{Wang2019}. This was achieved  without any external magnetic field and at room temperature as being highly relevant for applications. Furthermore, the absence of net magnetization in AFI forbids any  stray magnetic fields which makes such materials largely insensitive to perturbations by externally applied magnetic fields or those from neighboring layers~\cite{Jungfleisch2018,Baltz2018,Jungwirth2016,Zelezny2018}. Since insertion of a normal metal (NM) layer, such as Cu of thickness $\simeq 6$~nm, between NiO and Py layers did not substantially impede MSTT-driven magnetization switching of FM-analyzer, it was concluded~\cite{Wang2019} that direct exchange coupling between NiO and Py layers is not essential. Instead, one can conjecture that magnonic spin current is transmuted~\cite{Suresh2020,Bauer2011} at the AFI/NM interface into electronic spin current which then exerts conventional ESTT on the magnetization of the FM-analyzer. 

\begin{figure}
	\includegraphics[width =\linewidth]{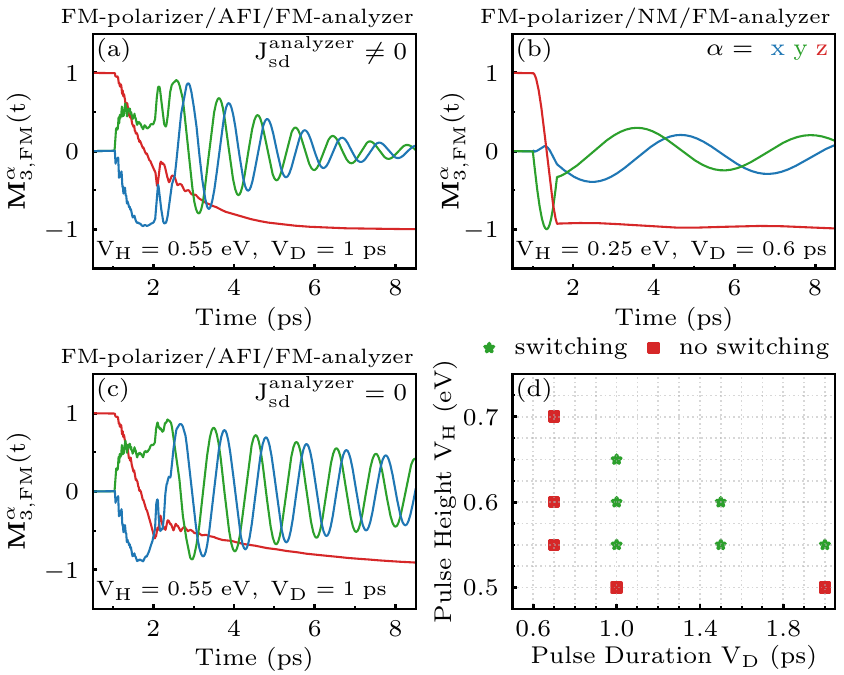}
	\caption{Time dependence of the last LMM, $\mathbf{M}_{3,\mathrm{FM}}(t)$ [Fig.~\ref{fig:fig1}], within the FM-analyzer due to current-pulse-initiated MSTT and/or ESTT in: (a),(c) FM-polarizer/AFI/FM-analyzer setup of Fig.~\ref{fig:fig1}(a); and (b) FM-polarizer/NM/FM-analyzer setup of Fig.~\ref{fig:fig1}(b). In panel (a) $J_{sd} \neq 0$ within the FM-analyzer, so that both MSTT and ESTT can operate concurrently. In panel (c) $J_{sd} = 0$ within the FM-analyzer, so that only MSTT due to directly injected magnonic spin current from AFI layer is operative, while ESTT due to concurrently injected (via pumping) electronic spin current is artificially turned off [Eq.~\eqref{eq:stt}]. Panel (d) lists the height $V_\mathrm{H}$ and the duration $V_\mathrm{D}$ of square voltage pulses in FM-polarizer/AFI/FM-analyzer setup with $J_{sd} \neq 0$ which cause either {\em no} switching (red squares), or switching (green stars) of LMMs  \mbox{$M_i^z=1 \mapsto M_i^z=-1$} as signified by red curve in panel (a) [note that one of the green stars corresponds to the voltage pulse employed in panel (a)].}
	\label{fig:fig2}
\end{figure}

Motivated by the experiments of Ref.~\cite{Wang2019}, we study MSTT in FM-polarizer/AFI/FM-analyzer setup illustrated in Fig.~\ref{fig:fig1}(a). We also  compare its efficiency with conventional ESTT in standard FM-polarizer/NM/FM-analyzer spin valve setup in Fig.~\ref{fig:fig1}(b). For this purpose, we employ recently developed multiscale and numerically exact {\em quantum-classical} framework~\cite{Petrovic2018,Bajpai2019a,Petrovic2019,Bostrom2019}. It combines  time-dependent nonequilibrium Green functions (TDNEGF)~\cite{Stefanucci2013,Gaury2014} description of conduction electrons out of equilibrium in open quantum systems, such as those illustrated in Fig.~\ref{fig:fig1} where the left (L) and right (R) macroscopic particle reservoirs make them open, with the Landau-Lifshitz-Gilbert (LLG) equation~\cite{Kim2010,Evans2014,Berkov2008} describing classical time evolution of LMMs. 

The classical treatment of LMMs, whose orientation is specified by unit vectors $\mathbf{M}_i(t)$,  is justified~\cite{Wieser2015} in the limit of large localized spins $S \rightarrow \infty$ and $\hbar \rightarrow 0$ (while $S \times \hbar  \rightarrow 1$), as well as in the  absence of entanglement~\cite{Mondal2019} between quantum states of individual LMMs. The latter condition is expected to be satisfied at room temperature (otherwise, since NiO is actually a strongly correlated insulator~\cite{Karolak2010}, we can expect its state to be highly entangled at low temperatures). We note that LLG description of the dynamics of local magnetization also appears in classical micromagnetics~\cite{Kim2010,Berkov2008}. But there $\mathbf{M}_i$ describe magnetization of a small volume of space, typically (2--10 nm)$^3$, rather than of individual atoms~\cite{Evans2014} that we have to assume in order to couple classical dynamics of $\mathbf{M}_i(t)$ to TDNEGF calculations where electrons hop from atom to atom. Furthermore, despite the ubiquity of micromagnetic simulations, they cannot~\cite{Evans2014} properly simulate  antiferromagnets or ferrimagnets whose intrinsic magnetization direction varies strongly on the atomic scale.

\begin{figure}
	\includegraphics[width =\linewidth]{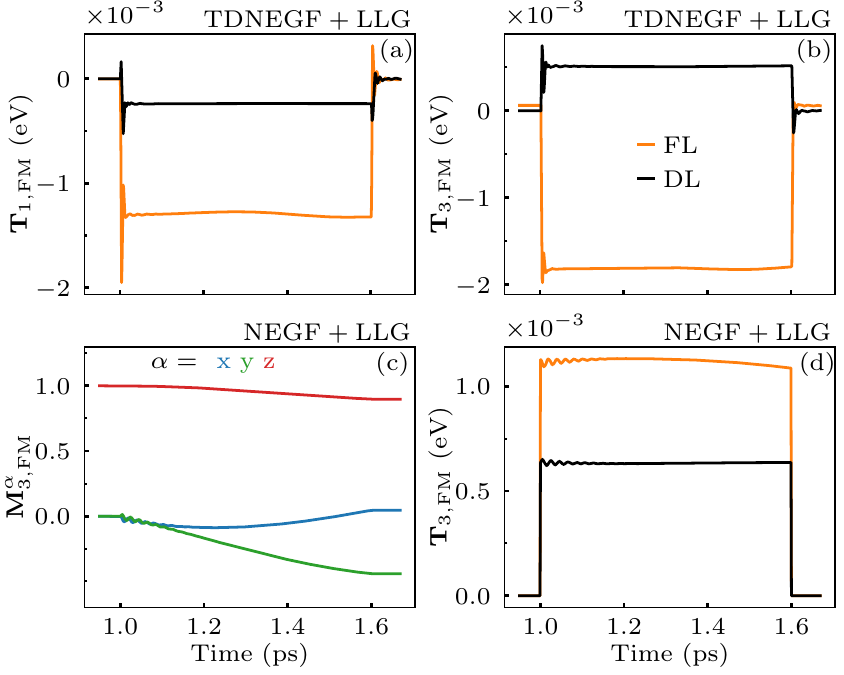}
	\caption{TDNEGF+LLG-computed time dependence of FL and DL components~\cite{Ralph2008} of ESTT vector $\mathbf{T}_i(t)$ [Eq.~\eqref{eq:stt}] on: (a) the first LMM of FM-analyzer; and (b) the last LMM of FM-analyzer within FM-polarizer/NM/FM-analyzer spin valve in Fig.~\ref{fig:fig1}(a). Panels  (c) and (d)  show steady-state-NEGF+LLG~\cite{Salahuddin2006,Lu2013,Ellis2017} computed time-dependence of  the last LMM [compare with Fig.~\ref{fig:fig2}(b)] of FM-analyzer and FL and DL components of ESTT vector [compare with panel (b)] on the last LMM of FM-analyzer, respectively. The  height \mbox{$V_\mathrm{H}=0.25$ V} and the duration \mbox{$V_\mathrm{D}= 0.6$ ps} of bias voltage square pulse employed  is the same as in 
	Fig.~\ref{fig:fig2}(b).}
	\label{fig:fig0}
\end{figure}

Our principal results for time evolution of LMMs and magnonic and electronic spin currents acting on them are summarized by Figs.~\ref{fig:fig2}--\ref{fig:fig6}, as well as their animations in embedded   Videos~\ref{vid:video1}--\ref{vid:video3}. The paper is organized as follows. In Sec.~\ref{sec:method} we introduce classical Hamiltonian for LMMs and quantum Hamiltonian for conduction electrons, where we also explain self-consistent coupling of  LLG and nonequilibrium density matrix (DM) calculations for classical and quantum dynamics, respectively. We warm up  by looking first in Sec.~\ref{sec:estt} at ESTT induced reversal of LMMs within the FM-analyzer of a conventional FM-polarizer/NM/FM-analyzer spin valve. This allows us to use such familiar case as a reference point, with new insights gained from TDNEGF+LLG approach when compared to standardly employed classical micromagnetic~\cite{Berkov2008} or previously developed  steady-state-NEGF+LLG approach~\cite{Salahuddin2006,Lu2013,Ellis2017}. In  Sec.~\ref{sec:mstt} we examine the interplay of MSTT and ESTT within the FM-analyzer of FM-polarizer/AFI/FM-analyzer, explicitly demonstrating that MSTT dominates and is required for magnetization switching [Fig.~\ref{fig:fig2}(c) and Video~\ref{vid:video2}], as the central subject of the paper. We conclude in Sec.~\ref{sec:conclusions}. An additional Appendix~\ref{sec:appendixa} provides TDNEGF-based derivation of continuity Eq.~\eqref{eq:spinconserve} connecting electronic spin density, spin currents and ESTT. 

\begin{video}
	\includegraphics[width=\linewidth]{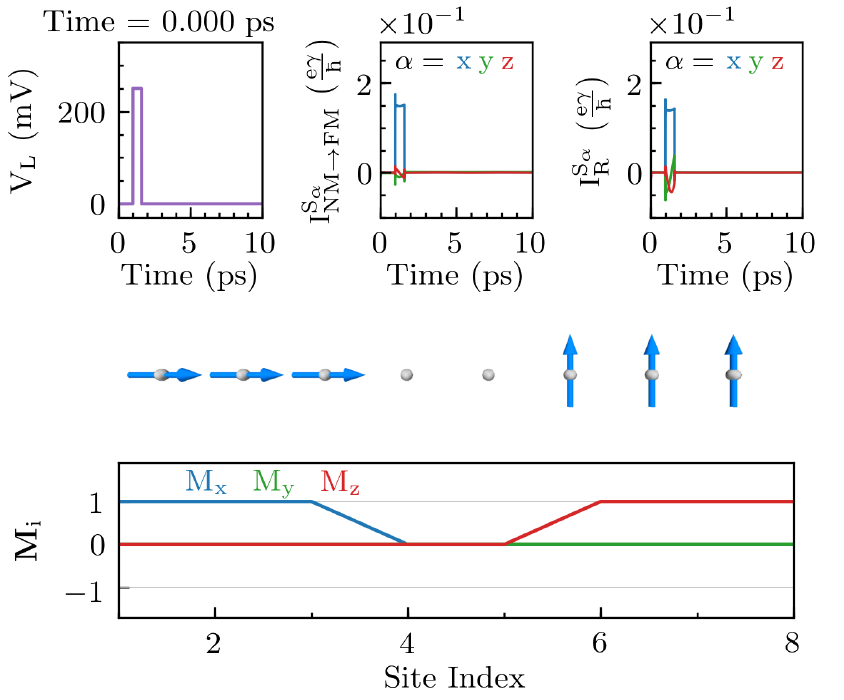}
	\setfloatlink{https://wiki.physics.udel.edu/wiki_qttg/images/4/4f/Video1.mp4}
	\caption{Animation of all $\mathbf{M}_i(t)$, including $\mathbf{M}_3(t)$ from Fig.~\ref{fig:fig2}(b), within FM-analyzer of FM-polarizer/NM/FM-analyzer spin valve setup in Fig.~\ref{fig:fig1}(b). Also animated are  time-dependences of: square voltage pulse $V_\mathrm{L}(t)$ applied	to the left NM lead; spin current $I_\mathrm{NM \rightarrow FM}^{S_\alpha}(t)$ from Fig.~\ref{fig:fig3}(e)--(g); and spin current $I_\mathrm{R}^{S_\alpha}(t)$ outflowing into the right NM lead.}
	\label{vid:video1}
\end{video}

\section{Classical and quantum Hamiltonian models and TDNEGF+LLG methodology} \label{sec:method}

Within the FM-polarizer in Fig.~\ref{fig:fig1}, we fix unit vectors $\mathbf{M}_i$ at each site $i$ of a one-dimensional (1D) tight-binding (TB) chain to point (and do not change in time) along the chain itself, i.e., along the $x$-axis as the direction of current flow. The classical Hamiltonian for the rest of LMMs in Fig.~\ref{fig:fig1},  which are allowed to evolve in time,  is given by
\begin{eqnarray}
\label{eq:classH}
{\mathcal H} & = &  J_\mathrm{AFI} \sum_{\langle ij \rangle \in \mathrm{AFI}}
\mathbf{M}_{i}\cdot \mathbf{M}_{j} - J_\mathrm{FM} \sum_{\langle ij \rangle \in \mathrm{FM}}
\mathbf{M}_{i}\cdot \mathbf{M}_{j} \nonumber \\
&& -J_\mathrm{AFI-FM} \mathbf{M}_{N,\mathrm{AFI}} \cdot \mathbf{M}_\mathrm{1,FM}  - J_{sd}\sum_{i} \langle \hat{\mathbf{s}}_i \rangle^\mathrm{CD}(t) \cdot
\mathbf{M}_{i} \nonumber \\ 
&& - K\sum_{i}
{\left(M_{i}^{z}\right)}^2. 
\end{eqnarray}
Here \mbox{$J_\mathrm{AFI} = 0.1\ {\rm eV}$} is the Heisenberg exchange coupling between the nearest-neighbor LMMs of AFI layer; \mbox{$J_\mathrm{FM} = 0.1\ {\rm eV}$} is the exchange coupling between the nearest-neighbor LMMs of FM-analyzer; \mbox{$J_\mathrm{AFI-FM}=0.1$ eV}, or \mbox{$J_\mathrm{AFI-FM}=0$} in some setups,  is the exchange coupling between the last LMM $\mathbf{M}_{N,\mathrm{AFI}}$ of AFI layer and the first LMM $\mathbf{M}_{1,\mathrm{FM}}$ of FM-analyzer; and magnetic anisotropy is specified by \mbox{$K = 0.00025$ eV} which selects the $z$-axis as the easy axis. In the case of spin valve in Fig.~\ref{fig:fig1}(b) we use the same Hamiltonian in Eq.~\eqref{eq:classH} but without AFI layer. The interaction of classical LMMs and current-driven (CD) part~\cite{Petrovic2018,Ellis2017} of nonequilibrium electronic spin density $\langle \hat{\mathbf{s}}_i \rangle^\mathrm{CD}(t)$ is described by $s$-$d$ exchange coupling of strength \mbox{$J_{sd}=0.1$ eV}, as measured experimentally~\cite{Cooper1967}. 

The classical dynamics of $\mathbf{M}_i(t)$ is obtained by solving a system of coupled LLG equations~\cite{Kim2010,Evans2014,Berkov2008}
\begin{equation}\label{eq:llg}
\frac{\partial\mathbf{M}_{i}}{\partial t} =
-\frac{g}{1 + {\lambda}^{2}}
\left[
\mathbf{M}_{i} \times \mathbf{B}^{\rm eff}_{i} 
+
\lambda \mathbf{M}_{i} \times 
\left(
\mathbf{M}_{i} \times \mathbf{B}^{\rm eff}_{i}
\right)
\right],
\end{equation}
using the Heun numerical scheme with projection to the unit sphere~\cite{Evans2014}. Here \mbox{$\mathbf{B}_{\rm eff}^{i} = - \frac{1}{\mu_M} \partial \mathcal{H} /\partial \mathbf{M}_{i}$}  is the effective magnetic field ($\mu_M$ is the magnitude of LMMs); $g$ is the gyromagnetic ratio; and the intrinsic Gilbert damping parameter $\lambda$ arises due to the well-established mechanism~\cite{Kambersky2007,Gilmore2007} combining spin-orbit coupling and electron-phonon interactions. We choose $\lambda = 0.005$ within the AFI layer and $\lambda = 0.05$ within the FM-analyzer.

\begin{figure*}
	\includegraphics[width =\linewidth]{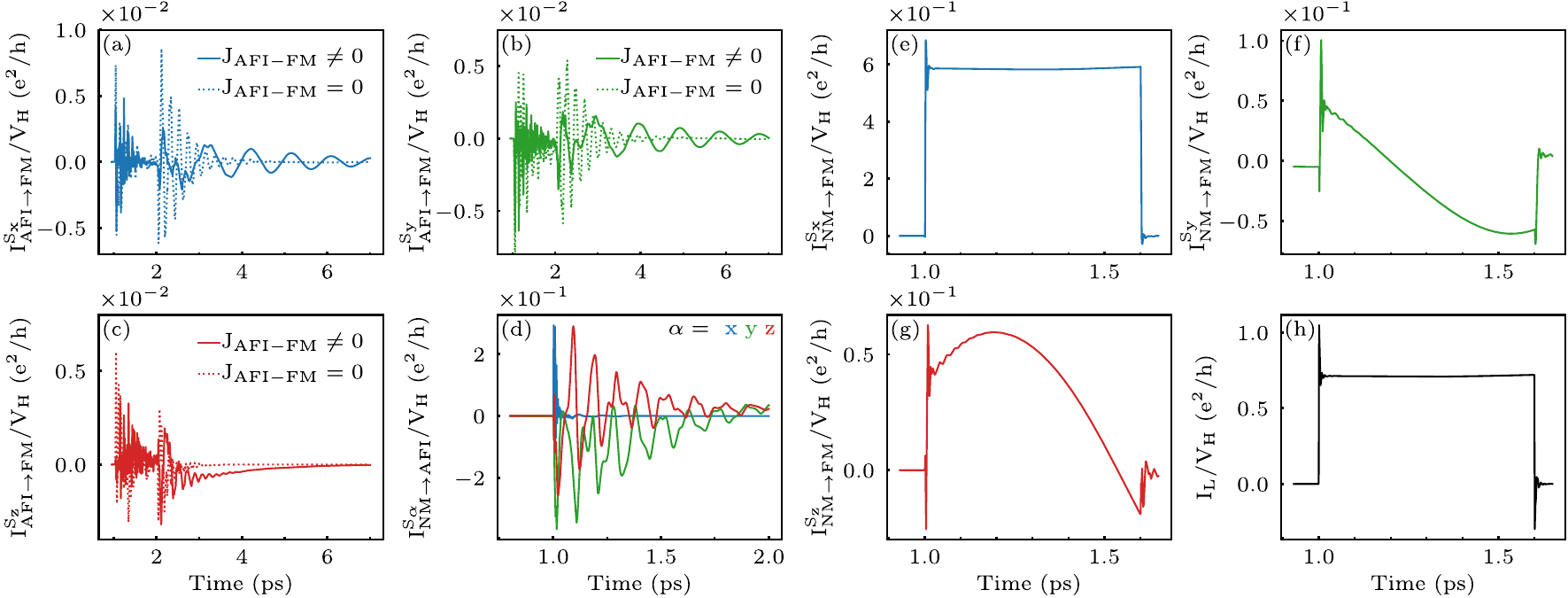}
	\caption{Time dependence of three components of bond {\em electronic} spin current [Eq.~\eqref{eq:bondspin}] between: (a)--(c) the last site of AFI and the first site of FM-analyzer denoted by $I^{S_\alpha}_\mathrm{AFI \rightarrow FM}(t)$ in Fig.~\ref{fig:fig1}(a), where dotted lines show the same spin current components injected into FM-analyzer after direct exchange coupling $J_\mathrm{AFI-FM}=0$ between AFI and FM-analyzer is set to zero; (e)--(g) the last site of NM interlayer and the first site of FM-analyzer denoted by $I^{S_\alpha}_\mathrm{NM \rightarrow FM}(t)$ in Fig.~\ref{fig:fig1}(b). Panel (d) shows time dependence of bond spin current  $I^{S_\alpha}_\mathrm{NM \rightarrow AFI}(t)$ impinging onto AFI and reflecting from its first LMM in the setup of Fig.~\ref{fig:fig1}(a), while panel (h) shows time dependence of charge current pulse injected from left NM lead by applying bias voltage square pulse of height \mbox{$V_\mathrm{H}=0.25$ V} and duration \mbox{$V_\mathrm{D}=0.6$ ps} in the setup of Fig.~\ref{fig:fig1}(b). The  height \mbox{$V_\mathrm{H}=0.55$ V} and the duration \mbox{$V_\mathrm{D}= 1.0$ ps} of bias voltage square pulse employed in panels (a)--(c) is the same as 
		in Fig.~\ref{fig:fig2}(a).}
	\label{fig:fig3}
\end{figure*}

\begin{video}
	\includegraphics[width=\linewidth]{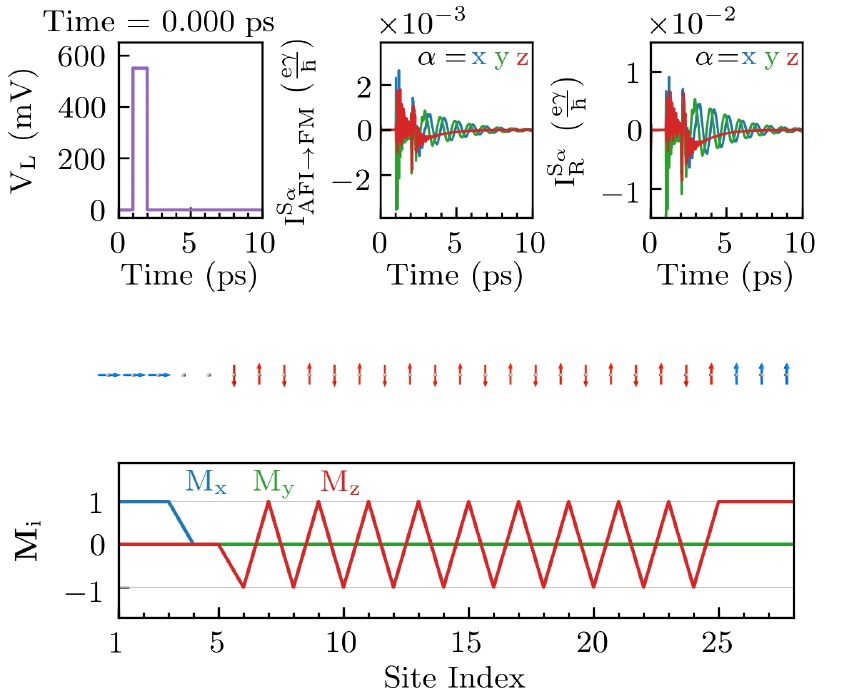}
	\setfloatlink{https://wiki.physics.udel.edu/wiki_qttg/images/3/36/Video2.mp4}
	\caption{Animation of all $\mathbf{M}_i(t)$, including $\mathbf{M}_3(t)$ from Fig.~\ref{fig:fig2}(a), within AFI layer and FM-analyzer  for {\em nonzero}  exchange coupling between AFI and FM,  $J_\mathrm{AFI-FM} \neq 0$, in FM-polarizer/AFI/FM-analyzer setup in Fig.~\ref{fig:fig1}(a). Also animated are time-dependences of: square voltage pulse $V_\mathrm{L}(t)$ applied to the left NM lead;  spin current $I_\mathrm{AFI \rightarrow FM}^{S_\alpha}(t)$ from Fig.~\ref{fig:fig3}(a)--(c); and spin current $I_\mathrm{R}^{S_\alpha}(t)$ outflowing into the right NM lead.}
	\label{vid:video2}
\end{video}

The conduction electron subsystem is modeled by a quantum Hamiltonian
\begin{equation} \label{eq:ham}
\hat{H} = - \sum_{\langle ij \rangle} \gamma_{ij} \hat{c}_{i}^\dagger\hat{c}_i - J_{sd} \sum_{i}\hat{c}_i^\dagger \hat{\bm \sigma} \cdot \bold{M}_i(t) \hat{c}_i,
\end{equation}
where the first term is a 1D TB model and the second term is the $s$-$d$ exchange coupling between LMMs and conduction electron spins described by the vector of the Pauli matrices $\mbox{$\hat{\bm  \sigma} = (\hat{\sigma}_x,\hat{\sigma}_y,\hat{\sigma}_z)$}$.  Here \mbox{$\hat{c}_i^\dagger = (\hat{c}_{i\uparrow}^\dagger,\hat{c}_{i\downarrow}^\dagger)$}  is a row vector containing operators $\hat{c}_{i\sigma}^\dagger$ which create an electron of spin $\sigma=\uparrow,\downarrow$ at the site $i$, and $\hat{c}_i$ is a column vector that contains the corresponding  annihilation operators. The semi-infinite NM leads attached to active region in Fig.~\ref{fig:fig1} are modeled by the first term alone in Eq.~\eqref{eq:ham}.  The nearest-neighbor hopping is \mbox{$\gamma_{ij}=1$ eV} in the NM leads, NM interlayer, FM-polarizer and FM-analyzer in Fig.~\ref{fig:fig1}, as well as between AFI layer and neighboring metallic layers. Within the AFI layer hopping is zero, \mbox{$\gamma_{ij} \equiv 0$} as denoted by dotted lines between its sites in Fig.~\ref{fig:fig1}(a), so that {\em no}  electron can propagate across it. The Fermi energy of macroscopic reservoirs in equilibrium for both junctions in Fig.~\ref{fig:fig1}  is chosen as \mbox{$E_F=-1.6$~eV}, which ensures maximum spin polarization of electronic current injected from the left NM lead.

\begin{video}
	\includegraphics[width=\linewidth]{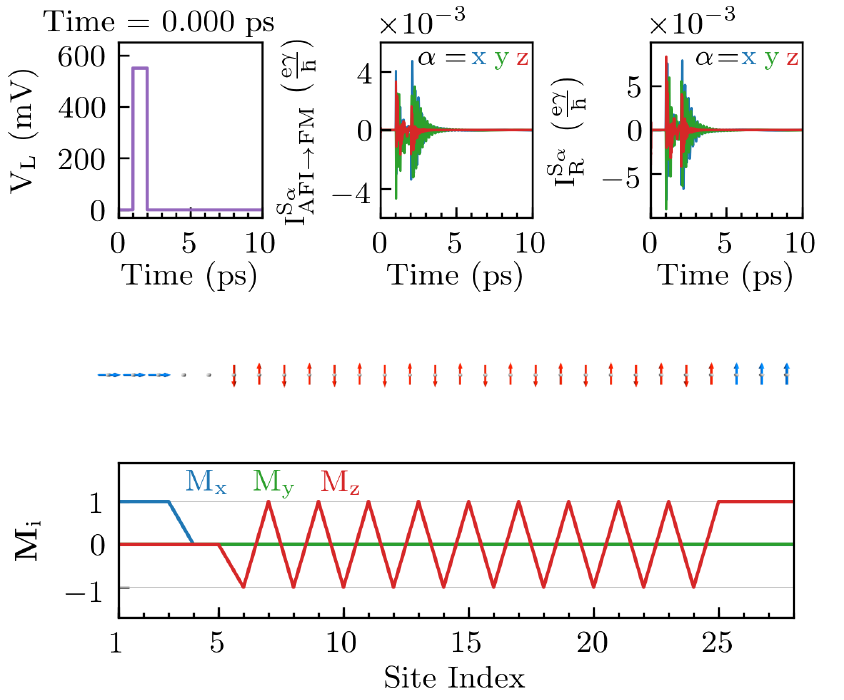}
	\setfloatlink{https://wiki.physics.udel.edu/wiki_qttg/images/7/7c/Video3.mp4}
	\caption{Same as Video~\ref{vid:video2} but for {\em zero}  exchange coupling between AFI layer and FM-analyzer,  $J_\mathrm{AFI-FM} = 0$.}
	\label{vid:video3}
\end{video}

The quantum dynamics of the electrons is described by solving a matrix integro-differential equation for time dependence of the nonequilibrium DM~\cite{Croy2009,Popescu2016,Petrovic2018}
\begin{equation}\label{eq:dm}
i\hbar \frac{d {\bm \rho}_\mathrm{neq}}{dt} = [\mathbf{H}, {\bm \rho}_\mathrm{neq}] + i \sum_{p=\mathrm{L,R}} [{\bm \Pi}_p(t) + {\bm \Pi}_p^\dagger(t)].
\end{equation}
This can be viewed as the exact master equation for an open finite-size quantum system, described by $\hat{H}$ and its matrix representation $\mathbf{H}$, that is attached (via semi-infinite NM leads) to macroscopic reservoirs. The matrices ${\bm \rho}_\mathrm{neq}$ and ${\bm \Pi}_p$ are expressed in terms of TDNEGFs~\cite{Stefanucci2013} and/or integrals over them, as elaborated in Refs.~\cite{Croy2009,Popescu2016}. The nonequilibrium electronic spin density is given by
\begin{equation}\label{eq:neq_spd}
\langle \hat{\mathbf{s}}_i\rangle^\mathrm{neq}(t) = \mathrm{Tr} \big\{ \bm{\rho}_\mathrm{neq}(t) |i\rangle\langle i| \otimes \hat{\bm{\sigma}} \big\},
\end{equation}
while its CD part~\cite{Petrovic2018,Ellis2017}
\begin{equation}\label{eq:cdspindensity}
\langle \hat{\mathbf{s}}_i \rangle^\mathrm{CD}(t) = \mathrm{Tr} \, \big\{ ( {\bm \rho}_\mathrm{neq}(t)-{\bm \rho}^\mathrm{eq}_t) |i\rangle \langle i| \otimes \hat{{\bm \sigma}} \big\},
\end{equation} 
appears in the classical Hamiltonian in Eq.~\eqref{eq:classH}. Therefore, it {\em generates} ESTT on each LMM
\begin{equation}\label{eq:stt}
\mathbf{T}_i(t)=J_{sd}\langle \hat{\mathbf{s}}_i \rangle^\mathrm{CD}(t) \times \mathbf{M}_i(t),
\end{equation}
within the LLG Eq.~\eqref{eq:llg} via $\mathbf{B}^{\rm eff}_{i}$. Here ${\bm \rho}^\mathrm{eq}_t$ is the grand canonical equilibrium DM~\cite{Petrovic2018} for instantaneous configuration of $\mathbf{M}_i(t)$ at time $t$, so that `adiabatic electronic spin density'~\cite{Stahl2017,Bajpai2020} 
\begin{equation}\label{eq:eqb_spd}
\langle \hat{\mathbf{s}}_i\rangle_t = \mathrm{Tr} \big\{ \bm{\rho}_t^\mathrm{eq} |i\rangle\langle i| \otimes \hat{\bm{\sigma}} \big\},
\end{equation}
determined by it  assumes $\partial  \mathbf{M}_i/\partial t= 0$ [subscript $t$ signifies  parametric dependence on time through $\mathbf{M}_i(t)$]. The ${\bm \Pi}_p$ matrices in Eq.~\eqref{eq:dm} yield the charge, \mbox{$I_p(t)=\frac{e}{\hbar} \mathrm{Tr} \{ {\bm \Pi}_p(t)\} $}, and the spin, \mbox{$I_p^{S_\alpha}(t)=\frac{e}{\hbar} \mathrm{Tr}\, \{ \hat{\sigma}_\alpha {\bm \Pi}_p(t) \} $}, currents flowing into the NM lead $p=\mathrm{L,R}$.  The local (bond) charge current~\cite{Nikolic2006} between sites $i$ and $j$ is computed as~\cite{Gaury2014} 
\begin{equation}\label{eq:bondcharge}
I_{i \rightarrow j}(t) =\frac{e}{i\hbar} \mathrm{Tr}_\mathrm{spin} \, \left\{ \gamma_{ji} \rho_\mathrm{CD}^{ij}(t) - \gamma_{ij}\rho_\mathrm{CD}^{ji}(t) \right\}, 
\end{equation}
and the local spin currents are given by 
\begin{equation}\label{eq:bondspin}
I_{i \rightarrow j}^{S_\alpha}(t) = \frac{e}{i\hbar} \mathrm{Tr}_\mathrm{spin} \, \left\{ \hat{\sigma}_\alpha \left( \gamma_{ji} \rho_\mathrm{CD}^{ij}(t) - \gamma_{ij} \rho_\mathrm{CD}^{ji}(t) \right) \right\}. 
\end{equation}
Here the CD part of the nonequilibrium  DM is obtained as $\rho_\mathrm{CD}^{ij}(t) = \rho_\mathrm{neq}^{ij}(t) - \rho^{\mathrm{eq},ij}_t$, and the trace is performed only in the spin factor space of total electronic Hilbert space $\mathcal{H} = \mathcal{H}_\mathrm{orbital} \otimes \mathcal{H}_\mathrm{spin}$. In our convention, {\em positive} value of any lead or bond current defined above means that charge or spin is flowing along the $+x$-axis. 

In the TDNEGF+LLG framework~\cite{Petrovic2018,Bajpai2019a,Petrovic2019}, we first solve for $\langle \hat{\mathbf{s}}\rangle^\mathrm{CD}_i(t)$ using Eqs.~\eqref{eq:dm} and ~\eqref{eq:cdspindensity}, which is then fed into Eq.~\eqref{eq:llg} to propagate LMMs $\mathbf{M}_i(t)$ in the next time step. These updated $\mathbf{M}_i(t)$ classical vectors are then fed back into the quantum Hamiltonian of the conduction electron subsystem in Eq.~\eqref{eq:ham} and DM in Eq.~\eqref{eq:dm} is updated. The active region in Fig.~\ref{fig:fig1} is disconnected from the NM leads at \mbox{$t=0$}; then we connect them through time evolution over a period of 1 ps during which ${\bm \rho}_\mathrm{neq}(t) \rightarrow {\bm \rho}^\mathrm{eq}$, so that all transient currents die out by \mbox{$t_0=1$ ps}; finally at \mbox{$t_0=1$ ps}, for {\em all} junctions in Figs.~\ref{fig:fig2}--\ref{fig:fig6} and Videos~\ref{vid:video1}--\ref{vid:video3}, square voltage pulses of various durations $V_\mathrm{D}$ and heights $V_\mathrm{H}$ are applied to drive them out of equilibrium. The time step \mbox{$\delta t=0.1$ fs} is used for numerical stability of TDNEGF calculations, as well as in  LLG calculations, and recently developed TDNEGF algorithms scaling linearly~\cite{Gaury2014,Popescu2016} in the number of time steps are employed to reach $\sim$ ps or $\sim$ ns time scales of relevance to spintronic and magnonic phenomena.  Thus obtained time-dependences of $\mathbf{M}_i(t)$, $\langle \hat{\mathbf{s}}_i \rangle^\mathrm{CD}(t)$, $\mathbf{T}_i(t)$, $I_{p}(t)$, $I_p^{S_{\alpha}}(t)$, $I_{i \rightarrow j}(t)$ and $I_{i \rightarrow j}^{S_\alpha}(t)$ are {\em numerically exact}. 

\section{Results and discussion} \label{sec:results}

\subsection{Electronic spin currents and LMM dynamics in FM-polarizer/NM/FM-analyzer spin valve} \label{sec:estt}

As a warm-up, we first consider conventional ESTT in a standard FM-polarizer/NM/FM-analyzer spin valve setup employed in seminal spin torque experiments ~\cite{Tsoi2000,Katine2000}, as well as in the early development~\cite{Stiles2002,Wang2008b} of steady-state quantum transport theories of ESTT. Although we use 1D chain to model the spin valve in Fig.~\ref{fig:fig1}(b), this can  be easily converted into a three-dimensional (3D) junction with macroscopic cross section by assuming that chain is periodically repeated in the $y$- and $z$-directions and $k$-point sampled~\cite{Thygesen2005}. This means that our TDNEGF calculations would have to be repeated at each $(k_y,k_z)$ point~\cite{Wang2008b}. When $\langle \hat{\mathbf{s}}_i \rangle^\mathrm{CD}$ and $\mathbf{M}_i$ of FM-analyzer are noncollinear, $\langle \hat{\mathbf{s}}_i \rangle^\mathrm{CD}$ due to propagating states and the corresponding ESTT $\propto \langle \hat{\mathbf{s}}_i \rangle^\mathrm{CD} \times \mathbf{M}_i$   oscillate~\cite{Wang2008b} within the FM-analyzer of such 3D junction as a function of position and without decaying~\cite{Stiles2002,Wang2008b}. Nevertheless, the transverse (with respect to $\mathbf{M}_i$) component of $\langle \hat{\mathbf{s}}_i \rangle^\mathrm{CD}$ is brought to zero (typically within $\sim 1$ nm in realistic materials like Co or Ni~\cite{Wang2008b}) away from the normal-metal/FM-analyzer interface by averaging over all incoming propagating states with different transverse wavevectors $(k_y,k_z)$. This is due to the fact that   frequency of spatial oscillations rapidly changes for different $(k_y,k_z)$~\cite{Stiles2002,Wang2008b}.  The ESTT can also have smaller contribution from evanescent states, which decays exponentially in space when moving away from the NM/FM-analyzer interface~\cite{Stiles2002,Wang2008b}.

\begin{figure}
	\includegraphics[width =\linewidth]{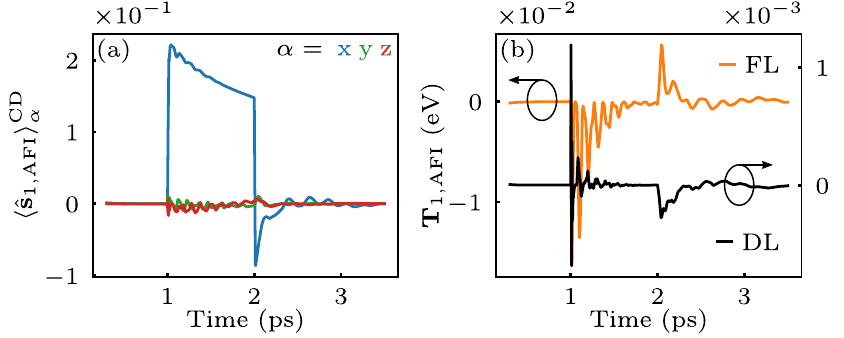}
	\caption{Time dependence of three Cartesian components of nonequilibrium spin density vector, $\langle \hat{\mathbf{s}}_i \rangle^\mathrm{CD}(t)$ [Eq.~\eqref{eq:cdspindensity}], at site $i \equiv \mathrm{1,AFI}$ of the first LMM of AFI within FM-polarizer/AFI/FM-analyzer junction in Fig.~\ref{fig:fig1}(a) studied in Fig.~\ref{fig:fig2}(a). The noncollinearity of this vector with  $\mathbf{M}_\mathrm{1,AFI}(t)$ leads to ESTT on the first LMM of AFI, $\mathbf{T}_\mathrm{1,AFI}(t)$ [Eq.~\eqref{eq:stt}], whose FL and DL components~\cite{Ralph2008} as a function of time are shown in panel (b). The  height \mbox{$V_\mathrm{H}=0.55$ V} and the duration \mbox{$V_\mathrm{D}= 1.0$ ps} of bias voltage square pulse employed  is the same as in Fig.~\ref{fig:fig2}(a).}
	\label{fig:fig4}
\end{figure}

Even though we effectively use only the $\Gamma$-point $(k_y,k_z)=(0,0)$ due to computational complexity of TDNEGF calculations, Fig.~\ref{fig:fig2}(b)  showing $\mathbf{M}_3(t)$ and accompanying Video~\ref{vid:video1} animating complete time evolution of all $\mathbf{M}_i(t)$ demonstrate that ESTT is deposited within  FM-analyzer in Fig.~\ref{fig:fig1}(b) to fully reverse all of its LMMs from positive to negative $z$-axis on the time scale comparable to voltage pulse duration $V_\mathrm{D}$. In widely-used classical micromagnetics~\cite{Berkov2008} ESTT has to be introduced as phenomenological term. More sophisticated  steady-state-NEGF+LLG simulations~\cite{Salahuddin2006,Lu2013,Ellis2017} compute ESTT microscopically from time-independent  quantum transport calculations, but they consider time as parameter rather than dynamical variable so that noncommutativity of electronic Hamiltonian at different times is lost. In contrast to this plausible approach, we can extract rigorously time dependence of standard~\cite{Ralph2008} field-like (FL) and damping-like (DL) components of ESTT 
\begin{equation}\label{eq:dlfl}
\mathbf{T}_i(t)= T_i^\mathrm{FL}(t) \mathbf{M}_i(t) \times \hat{\mathbf{x}} + T_i^\mathrm{DL}(t) \mathbf{M}_i(t) \times \left[\mathbf{M}_i(t) \times \hat{\mathbf{x}} \right], 
\end{equation}
from TDNEGF calculations, where $\hat{\mathbf{x}}$ is the unit vector along the $x$-axis. They are shown on the first and the last LMMs of FM-analyzer in Figs.~\ref{fig:fig0}(a) and ~\ref{fig:fig0}(b), respectively. 

It is also insightful to compare time-dependence of ESTT  from TDNEGF+LLG calculations in Figs.~\ref{fig:fig0}(b) to `time-dependence' from   steady-state-NEGF+LLG~\cite{Salahuddin2006,Lu2013,Ellis2017} calculations in Fig.~\ref{fig:fig0}(d)---this reveals incorrect magnitude of torque and/or its sign 
in the latter case. Furthermore, there is a qualitative difference since steady-state-NEGF+LLG simulations predict incorrectly no reversal of LMMs of FM-analyzer in Fig.~\ref{fig:fig0}(c), in contrast to TDNEGF+LLG simulations in Fig.~\ref{fig:fig2}(b). The discrepancy stems from the fact that steady-state-NEGF+LLG approach assumes electrons responding instantaneously to time-dependent potential introduced by $\mathbf{M}_i(t)$. Technically, this  means that only the lowest (adiabatic) term of the full nonadiabatic expansion of TDNEGF~\cite{Bajpai2019a,Mahfouzi2016,Bode2012} is included in steady-state-NEGF+LLG approach. Thus, this discrepancy demonstrates the importance of taking into account back-action~\cite{Bajpai2019a,Bajpai2020,Zhang2004,Zhang2009,Lee2013a,Sayad2015} of conduction electrons onto LMMs, as electrons are driven out of equilibrium both by the voltage pulse and the dynamics of $\mathbf{M}_i(t)$.  

\begin{figure}
	\includegraphics[width =\linewidth]{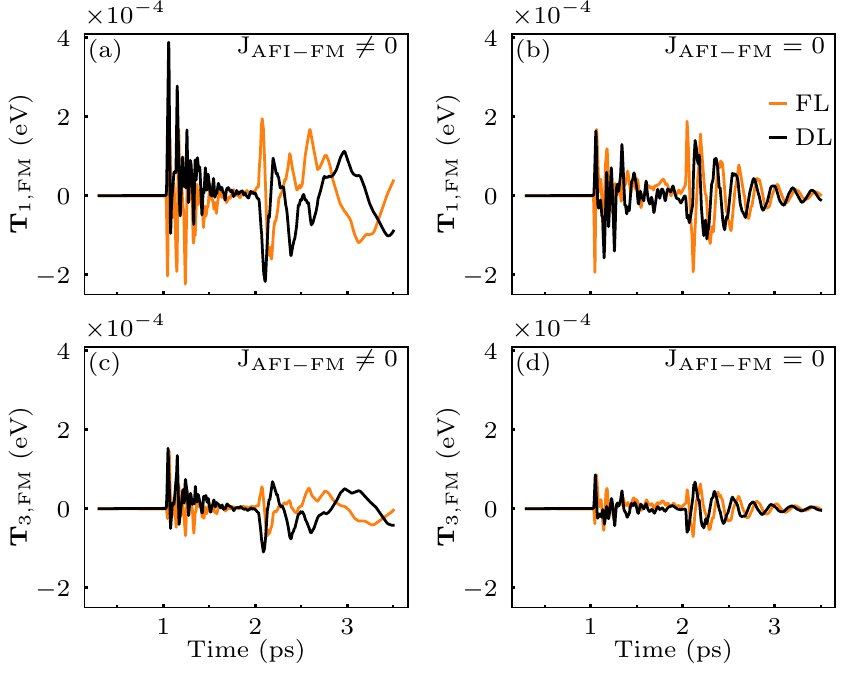}
	\caption{Time dependence of FL and DL components~\cite{Ralph2008} of ESTT vector $\mathbf{T}_\mathrm{i}(t)$ [Eq.~\eqref{eq:stt}] on: (a),(c) first LMM of FM-analyzer; and (b),(d) last LMM of FM-analyzer within FM-polarizer/AFI/FM-analyzer junction in Fig.~\ref{fig:fig1}(a).  The  height \mbox{$V_\mathrm{H}=0.55$ V} and the duration \mbox{$V_\mathrm{D}= 1.0$ ps} of bias voltage square pulse employed  is the same as in Fig.~\ref{fig:fig2}(a). In panels (a),(c) AFI layer and FM-analyzer are directly exchange coupled, $J_\mathrm{AFI-FM} \neq 0$, while in panels (b),(d) they are decoupled, $J_\mathrm{AFI-FM}=0$.}
	\label{fig:fig5}
\end{figure}

The time-dependence of injected unpolarized charge current pulse in the left NM lead $I_\mathrm{L}(t)$ is shown in Fig.~\ref{fig:fig3}(h), while three components of bond spin current $I^{S_\alpha}_\mathrm{NM \rightarrow FM}(t)$  impinging on the FM-analyzer to generate ESTT [Fig.~\ref{fig:fig0}] on its LMMs are plotted in Fig.~\ref{fig:fig3}(e)--(g). Besides expected \mbox{$I^{S_x}_\mathrm{NM \rightarrow FM}(t) \neq 0$} [Fig.~\ref{fig:fig3}(e)] component due to FM-polarizer with all of its LMMs pointing along the $x$-axis, there are also an order of magnitude smaller $I^{S_y}_\mathrm{NM \rightarrow FM}(t) \neq 0$ [Fig.~\ref{fig:fig3}(f)] and  $I^{S_z}_\mathrm{NM \rightarrow FM}(t) \neq 0$ [Fig.~\ref{fig:fig3}(g)]. This is attributed to electronic spin reflection and rotation at the NM/FM-analyzer interface, so that bond spin current $I^{S_\alpha}_\mathrm{NM \rightarrow FM}(t)$ is superposition of incoming current whose spins are polarized by FM-polarizer along the $x$-axis and reflected spin currents polarized in the other two directions. This explanation is confirmed by noticing that  $I^{S_y}_\mathrm{NM \rightarrow FM}(t)$ [Fig.~\ref{fig:fig3}(f)] and $I^{S_z}_\mathrm{NM \rightarrow FM}(t)$ [Fig.~\ref{fig:fig3}(g)] turning negative in the course of their time evolution means that those currents flow backward toward the left NM lead in our convention for the sign of bond current.  

\subsection{Magnonic and electronic spin currents and LMM dynamics in FM-polarizer/AFI/FM-analyzer junction}\label{sec:mstt}

When unpolarized charge current pulse is injected from the left NM lead into FM-polarizer/AFI/FM-analyzer junction in Fig.~\ref{fig:fig1}(a), it gets spin-polarized along the $x$-axis and is subsequently fully reflected by AFI layer because of zero hopping (\mbox{$\gamma_{ij} \equiv 0$}) between its sites. In this process, the other two components of bond  spin current, $I_\mathrm{NM \rightarrow AFI}^{S_y} \neq 0 $ and $I_\mathrm{NM \rightarrow AFI}^{S_z} \neq 0$,  emerge in Fig.~\ref{fig:fig3}(d). Due to back and forth reflection between AFI and FM-polarizer, they will change direction in oscillatory fashion which is the meaning of sign change of $I_\mathrm{NM \rightarrow AFI}^{S_\alpha}(t)$ shown in Fig.~\ref{fig:fig3}(d). 

Note that in the experiment of Ref.~\cite{Wang2019}, the unpolarized charge current pulse is injected parallel to the TI/AFI interface    
and polarized by the TI metallic surface~\cite{Bansil2016} to have the largest component of nonequilibrium electronic spin density vector lying~\cite{Chang2015} in the plane of the interface. Thus, features in Fig.~\ref{fig:fig3}(d), where charge current pulse is injected perpendicular to NM/AFI interface, would not be directly pertinent to the experiment. Nevertheless, in both cases nonequilibrium electronic spin density $\langle \hat{\mathbf{s}}_\mathrm{1,AFI} \rangle^\mathrm{CD}(t)$ [Eq.~\eqref{eq:cdspindensity}] plotted in Fig.~\ref{fig:fig4}(a) is noncollinear to the first LMM of AFI, denoted by $\mathbf{M}_\mathrm{1,AFI}(t)$, which then leads to ESTT on it. The time dependence of standard~\cite{Ralph2008} FL and DL components of ESTT acting on the first LMM of AFI is shown in Fig.~\ref{fig:fig4}(b). The ESTT-driven dynamics of $\mathbf{M}_\mathrm{1,AFI}(t)$ then generates SW within AFI because of the exchange coupling between its LMMs [encoded by the first term on the right-hand side of Eq.~\eqref{eq:classH}]. This picture is also confirmed by Videos~\ref{vid:video2} and ~\ref{vid:video3} which animate $\mathbf{M}_i(t)$ for all 20  LMMs of the AFI layer. Thus, this is the same final outcome as in the experiment of Ref.~\cite{Wang2019}---SW propagation is initiated across AFI---independently of specific mechanism applied to one of the surface of AFI in order to trigger the SW across it.

\begin{figure}
	\includegraphics[width =\linewidth]{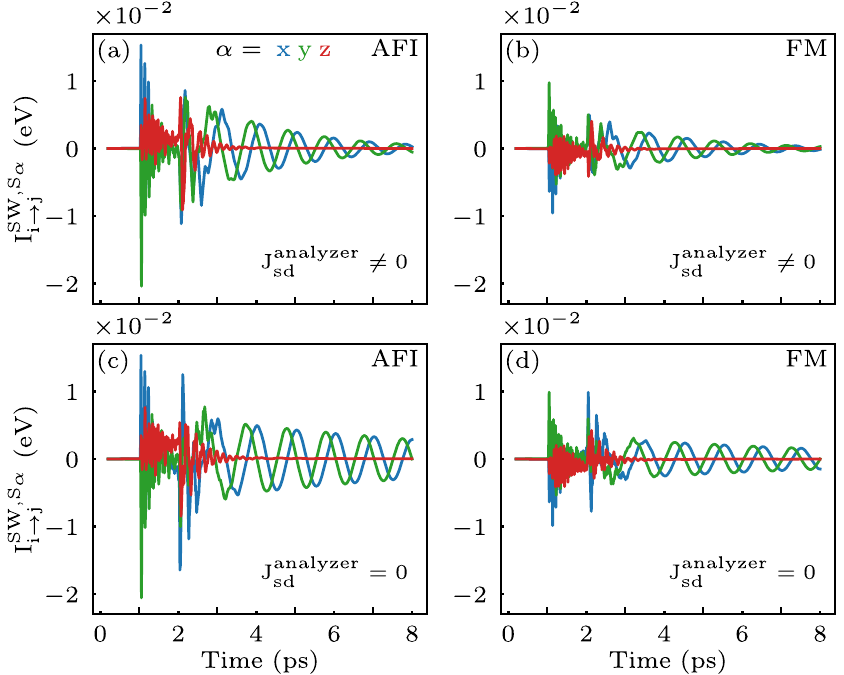}
	\caption{Time dependence of three components of bond {\em magnonic} spin current [Eq.~\eqref{eq:bondsw}] between: (a),(c) last two sites of AFI layer; and (b),(d) first two sites [$1$ and $2$ in Fig.~\ref{fig:fig1}(a)] of the FM-analyzer. The AFI layer and FM-analyzer are directly exchange coupled, $J_\mathrm{AFI-FM} \neq 0$, so that magnonic spin current can flow between them. The $s$-$d$ exchange coupling between conduction electron spins and LMMs is nonzero in (a) and (b), and set to zero within FM-analyzer in (c) and (d).  The  height \mbox{$V_\mathrm{H}=0.55$ V} and the duration \mbox{$V_\mathrm{D}= 1.0$ ps} of bias voltage square pulse employed in all panels is the same as in Fig.~\ref{fig:fig2}(a).}
	\label{fig:fig6}
\end{figure}

When  SW reaches the last LMM of AFI layer, it will initiate its dynamics. Since TB site of this last LMM of AFI layer is connected directly by nonzero hopping to TB site of LMM 1 within the FM-analyzer, as well as via $J_{sd} \neq 0$ to electronic spins within the FM-analyzer, its dynamics will pump electronic spin current~\cite{Tserkovnyak2005,Tatara2019,Tatara2017,Chen2009} into the FM-analyzer (as well as charge current because the left-right symmetry of the device is broken~\cite{FoaTorres2005,Chen2009,Bajpai2019}). The time-dependence of pumped bond spin current $I_\mathrm{AFI \rightarrow FM}^{S_\alpha}(t)$ between the last LMM of AFI layer and the first LMM of the FM-analyzer is plotted in Fig.~\ref{fig:fig3}(a)--(c) and animated in Videos~\ref{vid:video2} and ~\ref{vid:video3}. Additionally, Videos~\ref{vid:video2} and ~\ref{vid:video3} show pumped spin current $I_\mathrm{R}^{S_\alpha}(t)$ outflowing into the right NM lead. Recall that all of $I_\mathrm{AFI \rightarrow FM}^{S_\alpha}(t)$ and $I_\mathrm{R}^{S_\alpha}(t)$ {\em must} be generated by spin pumping because none of the originally injected spin-polarized current pulse can pass through the AFI layer. In the absence of spin-flips, due to spin-orbit coupling in the band structure or spin-orbit and/or magnetic impurities~\cite{Kohno2006}, the sum of $\mathbf{T}_i$ [Eq.~\eqref{eq:stt}] on each LMM is equal to the sum of absorbed spin current ~\cite{Stiles2002,Wang2008b} and the rate of change of total nonequilibrium electronic spin density within the FM-analyzer at each instant of time $t$
\begin{equation}\label{eq:spinconserve}
\bigg[ \sum_i \mathbf{T}_i(t) \bigg]_\alpha = \frac{\hbar}{2e} \left[ I_\mathrm{AFI \rightarrow FM}^{S_\alpha}(t)-I^{S_\alpha}_\mathrm{R}(t) \right ] +  \sum_{i} \frac{\hbar}{2} \frac{\partial\langle\hat{\mathrm{s}}^\alpha_i\rangle^\mathrm{neq}}{\partial t},  
\end{equation}
Appendix~\ref{sec:appendixa} provides rigorous TDNEGF-based derivation of such spin density continuity equation.

When direct exchange coupling between AFI and FM-analyzer is absent, $J_\mathrm{AFI-FM}=0$, spin current $I_\mathrm{AFI \rightarrow FM}^{S_\alpha}(t)$ can also be viewed as the result of transmutation~\cite{Suresh2020,Bauer2011} of magnonic spin-current into solely electronic spin current within the FM-analyzer layer via spin pumping by the last LMM of AFI layer. Such electronic spin current, plotted by dotted line in Fig.~\ref{fig:fig3}(a)--(c), turns out to be {\em insufficient} to reverse LMMs of FM-analyzer, as visualized by  Video~\ref{vid:video3}. According to Eq.~\eqref{eq:spinconserve}, this means that insufficient spin angular momentum is transferred to LMMs of the FM-analyzer. 

The importance of $J_\mathrm{AFI-FM} \neq  0$ in Fig.~\ref{fig:fig2}(a) and Video~\ref{vid:video2} for complete reversal of all LMMs within the 
FM-analyzer motivates us to further examine SW transmission into the FM-analyzer and thereby induced MSTT. The SW transmitted from AFI layer to FM-analyzer enhances injected electronic spin current $I_\mathrm{AFI \rightarrow FM}^{S_\alpha}(t)$ [solid lines in Fig.~\ref{fig:fig3}(a)--(c)] by SW-generated spin pumping~\cite{Suresh2020}. This then leads to larger ESTT on LMMs of FM-analyzer in Figs.~\ref{fig:fig5}(a) and ~\ref{fig:fig5}(c) than for the case when $J_\mathrm{AFI-FM} =  0$ in Figs.~\ref{fig:fig5}(b) and ~\ref{fig:fig5}(d). Nonetheless, we see that ESTT in Fig.~\ref{fig:fig5} due to pumped electronic spin current from AFI is an order of magnitude smaller than ESTT in spin valve in Fig.~\ref{fig:fig0} where electronic spin current pulse 
is directly injected from FM-polarizer into FM-analyzer. 

Since SW transmitted into FM-analyzer carries its own magnonic spin current, defined for continuous local magnetization $\mathbf{M}(x,t)$ as~\cite{Yan2011} 
\begin{equation} \label{eq:swyan}
\mathbf{j}^\mathrm{SW} \propto \mathbf{M} \times \frac{\partial \mathbf{M}}{\partial x}, 
\end{equation}
this means that {\em both} MSTT and ESTT act on LMMs of FM-analyzer  in Fig.~\ref{fig:fig2}(a) and Video~\ref{vid:video2}. In order to separate their individual contributions to total spin-transfer torque, we artificially turn off ESTT [Eq.~\eqref{eq:stt}] by setting $J_{sd} = 0$ within the FM-analyzer in Fig.~\ref{fig:fig2}(c). This reveals that LMMs still reverse, on slightly longer time scale [Fig.~\ref{fig:fig2}(c)], thereby confirming that MSTT is the {\em dominant mechanism} of magnetization switching in FM-polarizer/AFI/FM-analyzer junction. 

At first sight, this finding, together with Video~\ref{vid:video3} explicitly showing  inability of ESTT alone to reverse LMMs of FM-analyzer via magnonic-to-electronic spin current conversion at AFI/FM-analyzer interface when $J_\mathrm{AFI-FM} =  0$, differs from experimental conclusion~\cite{Wang2019}. In the experiment~\cite{Wang2019}, torque from AFI layer to FM-analyzer is slightly impeded by inserting $6$ nm of Cu in between them, which certainly suppresses direct exchange coupling $J_\mathrm{AFI-FM} = 0$ that is operative on \mbox{$\sim 0.1$ nm} length scale. However, to obtain switching of the magnetization of FM-analyzer with Cu layer inserted, the current density of injected pulse also had to be increased in the experiment (see Fig. 4 in Ref.~\cite{Wang2019}), which is compatible with our finding. 

The MSTT is driven by magnonic spin currents, which are visualized in Fig.~\ref{fig:fig6} by plotting time dependence of bond magnonic spin current~\cite{Schuetz2004} 
\begin{equation}\label{eq:bondsw}
(I_{i \rightarrow j}^{\mathrm{SW},S_x},I_{i \rightarrow j}^{\mathrm{SW},S_y},I_{i \rightarrow j}^{\mathrm{SW},S_z}) = J_{ij} \mathbf{M}_i \times \mathbf{M}_{j},
\end{equation}
that can also be viewed as the discretized version of Eq.~\eqref{eq:swyan}. This is plotted between sites $i$ and $j$, which are either last two sites of the AFI layer  [Figs.~\ref{fig:fig6}(a) and ~\ref{fig:fig6}(c)] or first two sites of the FM-analyzer [Figs.~\ref{fig:fig6}(b) and ~\ref{fig:fig6}(d)], so that $J_{ij} \equiv J_\mathrm{AFI}$ or  $J_{ij} \equiv J_\mathrm{FM}$, respectively. When $J_{sd} = 0$ is turned off within the FM-analyzer, bond magnonic spin current within the FM-analyzer decays on longer time scale in Figs.~\ref{fig:fig6}(c) and ~\ref{fig:fig6}(d) because 
of the elimination of a damping channel~\cite{Bajpai2019a,Tatara2019,Zhang2004} via conduction electrons that are otherwise (i.e., when $J_{sd} \neq 0$) driven out of equilibrium by SW dynamics of LMMs. Note that in Figs.~\ref{fig:fig6}(c) and ~\ref{fig:fig6}(d) we keep $J_{sd} \neq 0$ between the last LMM of AFI and electrons on the right side of it. So, its dynamics pumps the same electronic spin current as in Figs.~\ref{fig:fig6}(a) and ~\ref{fig:fig6}(b), but  such current does not interact with LMMs of the FM-analyzer. 

The MSTT alone in Fig.~\ref{fig:fig2}(c) or combined MSTT and ESTT in Fig.~\ref{fig:fig2}(a) show that they are less efficient, in the sense of requiring larger pulse height and its longer duration, than conventional ESTT-driven LMM switching  [Fig.~\ref{fig:fig2}(b)] in standard  FM-polarizer/normal-metal/FM-analyzer spin valves. Figure~\ref{fig:fig2}(d) summarizes different pulse heights and durations for the setup studied in Fig.~\ref{fig:fig2}(a) which are capable (green stars) to switch LMMs, or just initiate their dynamics without leading to full reversal (red squares).  The higher efficiency of ESTT-driven LMM switching in standard  FM-polarizer/normal-metal/FM-analyzer spin valves is not surprising in the sense that the same electronic current pulse injected from the left NM lead and spin-polarized by FM-polarizer layer in both setups in Fig.~\ref{fig:fig1} will encounter more interfaces in Fig.~\ref{fig:fig1}(a) at which it is  reflected and/or converted into magnonic current pulse. Both of this processes lead to reduced net spin angular momentum that can be deposited into the FM-analyzer via spin-transfer torque. Nevertheless, although FM-polarizer/AFI/FM-analyzer junction [Fig.~\ref{fig:fig1}(a)] is not efficient for applications (as is the case also of  Bi$_2$Se$_3$/NiO/Py heterostructure in experiment of Ref.~\cite{Wang2019}), it offers a setting to controllably study properties of MSTT and how to optimize it toward all-magnon devices. 

\section{Conclusions} \label{sec:conclusions}
	
In conclusion, using quantum-classical simulations based on recently developed TDNEGF+LLG framework~\cite{Petrovic2018,Bajpai2019a,Petrovic2019,Bostrom2019}, 
we provide microscopic (i.e., Hamiltonian-based) understanding of how MSTT and ESTT processes are initiated by magnonic and electronic spin currents, respectively, while these currents interconvert at different interfaces of FM-polarizer/AFI/FM-analyzer setup inspired by very recent experiment~\cite{Wang2019}. To assist such understanding, we provide Videos~\ref{vid:video2} and ~\ref{vid:video3} revealing how electronic spin current, after spin-polarization by the FM-polarizer in Fig.~\ref{fig:fig1}(a), is reflected at the  FM-polarizer/AFI interface to ignite magnonic spin current across the AFI layer since no electrons can move through it. Upon reaching the opposite edge of AFI layer, magnonic spin current initiates pumping of electronic spin current into the FM-analyzer by the dynamics of edge LMMs of AFI, while concurrently transmitting  into the FM-analyzer if direct exchange coupling is present between the LMMs of AFI and FM-analyzer. This means that, in general, FM-analyzer {\em receives combined MSTT and ESTT} which can completely reverse the direction of all of its LMMs. Akin to experiments (see Fig. 3C in Ref.~\cite{Wang2019}), switching by ESTT alone in FM-polarizer/NM/FM-analyzer setup with AFI layer removed is still more efficient, by requiring shorter voltage pulses of lower height [Fig.~\ref{fig:fig2}], than combined MSTT and ESTT in FM-polarizer/AFI/FM-analyzer setup. By artificially turning off ESTT [Fig.~\ref{fig:fig2}(c)], we demonstrate that MSTT, due to magnonic spin current flowing directly from AFI to FM-analyzer, is actually the dominant mechanism of LMMs reversal within the FM-analyzer. Despite smaller efficiency, MSTT in FM-polarizer/AFI/FM-analyzer junction offers a playground for detailed theoretical understanding of necessary vs. unnecessary phenomena for the development of {\em all-magnon-driven} magnetization switching without involving any electronic parts.
 
\begin{acknowledgments}
A.~S., U.B., M.~D.~P. and B.~K.~N. were supported by the US National Science Foundation (NSF) under Grant No. ECCS 1922689. H.~Y. was supported by SpOT-LITE programme (A*STAR Grant no. 18A6b0057) through RIE2020 funds from Singapore and Samsung Electronics’ University R\&D program (Exotic SOT materials/SOT characterization). This research was initiated during {\em Spin and Heat Transport in Quantum and Topological Materials} program at KITP Santa Barbara, which is supported under NSF Grant No. PHY-1748958.
\end{acknowledgments}

\bigskip

\appendix

\section{Derivation of spin density continuity Eq.~\eqref{eq:spinconserve}}\label{sec:appendixa}

The spin density continuity Eq.~\eqref{eq:spinconserve} can be rigorously derived within the TDNEGF+LLG formalism. The rate of change of the nonequilibrium electronic spin density $\langle\hat{\mathbf{s}}_i\rangle^\mathrm{neq}(t)$ defined in Eq.~\eqref{eq:neq_spd} is given by
\begin{widetext} 
\begin{equation}\label{eq:spd}
    \frac{\hbar}{2}\frac{\partial\langle\hat{\mathrm{s}}^\alpha_i\rangle^\mathrm{neq}}{\partial t} 
       = \frac{\hbar}{2}
     \mathrm{Tr}\left\{ \frac{\partial \bm{\rho}_\mathrm{neq}}{\partial t} \hat{\sigma}^\alpha_i \right\} 
       =  
    \frac{1}{2i} \mathrm{Tr} \left\{ \bm{\rho}_\mathrm{neq} [  \hat{\sigma}^\alpha_i, \bold{H}(t) ] \right\}
      +  \frac{1}{2}\sum_{p = \mathrm{L, R}}  \mathrm{Tr} \left\{ \left( \bm{\Pi}_p(t) + \bm{\Pi}^\dagger_p (t) \right)  \hat{\sigma}^\alpha_i \right\},
\end{equation}
where  Eq.~\eqref{eq:dm} was used in the last step and we employ notation $\hat{\sigma}_i^\alpha = |i\rangle\langle i| \otimes \hat{\sigma}_\alpha$. The first term in Eq.~\eqref{eq:spd} represents the full ESTT $\bold{T}_i^\mathrm{neq}(t)$ exerted on LMMs $\mathbf{M}_i(t)$
\begin{equation}\label{eq:torque_part}
\left[\bold{T}_i^\mathrm{neq}(t) \right]_\alpha = J_{sd} \left[   \langle \hat{\bm{\mathrm{s}}}_i \rangle^\mathrm{neq}(t) \times \bold{M}_i(t) \right]_\alpha = \frac{1}{2i}\mathrm{Tr} \big \{ \bm{\rho}_\mathrm{neq} [\hat{\sigma}^\alpha_i, \bold{H}(t)] \big \} 
  = J _{sd}\sum_{\beta\gamma} \epsilon^{\alpha\beta\gamma} M^\beta_i(t) \mathrm{Tr} \big \{ \bm{\rho}_\mathrm{neq}\hat{\sigma}^\gamma_i \big \},
\end{equation}
where $\epsilon^{\alpha\beta\gamma}$ is the Levi-Civita symbol. The full ESTT can be decomposed as 
\begin{equation}
\bold{T}^\mathrm{neq}_i(t) = \bold{T}_i(t) + \bold{T}_{i,t},
\end{equation}
where $\bold{T}_i(t)$ is the CD part [Eq.~\eqref{eq:stt}] of ESTT while $\bold{T}_{i,t} = J_{sd}  \langle \hat{\bold{s}}_i \rangle_t \times \bold{M}_i(t)$ is the adiabatic  part which originates due to always present~\cite{Stahl2017,Bajpai2020} noncollinearity  of LMMs $\mathbf{M}_i(t)$ and `adiabatic electronic spin density' $\langle \hat{\bold{s}}_i \rangle_t$ introduced in Eq.~\eqref{eq:eqb_spd}. The second term in Eq.~\eqref{eq:spd} can be recast as
\begin{equation}\label{eq:current_part}
\frac{1}{2}\sum_{p = \mathrm{L, R}}\mathrm{Tr}\left\{ \left( \bm{\Pi}_p(t) + \bm{\Pi}_p^\dagger(t) \right) \hat{\sigma}^\alpha _i \right\}  = \frac{\hbar}{2e}\left[ \mathcal{I}^{S_\alpha}_{i \to i+1}(t) - \mathcal{I}^{S_\alpha}_{i \to i-1}(t)\right],
\end{equation}
where  $\mathcal{I}_{i \to j}^{S_\alpha}(t)$ is the nonequilibrium local spin current flowing from site $i$ to $j$ that can be computed from Eq.~\eqref{eq:bondspin} by replacing $\rho_\mathrm{CD}^{ij}(t) \mapsto \rho_\mathrm{neq}^{ij}(t)$. Additionally, $\mathcal{I}_{i \to j}^{S_\alpha}(t)$ can also be expressed as a sum of its CD contribution $I_{i \to j}^{S_\alpha}(t)$ [Eq.~\eqref{eq:bondspin}], and adiabatic contribution $I_{{i \to j},t}^{S_\alpha}$ which is obtained through  Eq.~\eqref{eq:bondspin} by replacing $\rho_\mathrm{CD}^{ij}(t) \mapsto \rho^{\mathrm{eq},ij}_t$
\begin{equation}\label{eq:tot_curr}
\mathcal{I}_{i \to j}^{S_\alpha}(t) = I_{i \to j}^{S_\alpha}(t) + I_{{i \to j},t}^{S_\alpha}.
\end{equation}
Inserting Eq.~\eqref{eq:torque_part}--\eqref{eq:tot_curr} into Eq.~\eqref{eq:spd} yields
\begin{equation}\label{eq:torq_curr_spd}
 \big[ \bold{T}_i(t) + \bold{T}_{i,t} \big]_\alpha 
   =
 \frac{\hbar}{2e} \left[ I^{S_\alpha}_{i \to i+1}(t) - I^{S_\alpha}_{i \to i-1}(t) \right] 
   +
 \frac{\hbar}{2e} \left[ I^{S_\alpha}_{i \to i+1,t} - I^{S_\alpha}_{i \to i-1,t} \right] 
   +   \frac{\hbar}{2} \frac{\partial\langle\hat{\mathrm{s}}^\alpha_i\rangle^\mathrm{neq}}{\partial t}.
\end{equation}
In the adiabatic limit (assuming $\partial \bold{M}_i/\partial t = 0$), we have $\bold{T}_i(t) \equiv 0$ and $I_{i\to j}^{S_\alpha}(t) \equiv 0$ while $\partial \langle\hat{\mathbf{s}}_i\rangle^\mathrm{neq}/\partial t = 0$. Therefore, Eq.~\eqref{eq:torq_curr_spd} furnishes an identity
\begin{equation}
\big[ \bold{T}_{i,t} \big]_\alpha \equiv \frac{\hbar}{2e} \left[ I^{S_\alpha}_{i \to i+1,t} - I^{S_\alpha}_{i \to i-1,t} \right],
\end{equation}
so that spin density continuity equation at an arbitrary site $i$ is given by
\begin{equation}\label{eq:onsite}
\big[ \bold{T}_i(t) \big]_\alpha = \frac{\hbar}{2e} \left[ I^{S_\alpha}_{i \to i+1}(t) - I^{S_\alpha}_{i \to i-1}(t)\right] + \frac{\hbar}{2} \frac{\partial\langle\hat{\mathrm{s}}^\alpha_i\rangle^\mathrm{neq}}{\partial t}.
\end{equation}
Finally, summing Eq.~\eqref{eq:onsite} over all sites $i$ within the FM-analyzer layer in Fig.~\ref{fig:fig1} leads to 
\begin{equation}
\bigg[ \sum_{i} \bold{T}_i(t) \bigg]_\alpha = \frac{\hbar}{2e} \left[ I^{S_\alpha}_\mathrm{AFI \to FM}(t) - I^{S_\alpha}_\mathrm{R}(t)\right] +  \sum_{i} \frac{\hbar}{2} \frac{\partial\langle\hat{\mathrm{s}}^\alpha_i\rangle^\mathrm{neq}}{\partial t},
\end{equation}
thereby completing the proof of Eq.~\eqref{eq:spinconserve}.
\end{widetext}



	

\end{document}